# Effect of annealing on the magnetic and superconducting properties of single-crystalline UCoGe

N.T. Huy[*], Y.K. Huang and A. de Visser

*Van der Waals - Zeeman Institute, University of Amsterdam,*
*Valckenierstraat 65, 1018 XE Amsterdam, The Netherlands*

**Abstract**

Single-crystals of the new ferromagnetic superconductor UCoGe have been grown. The quality of as-grown samples can be significantly improved by a heat-treatment procedure, which increases the residual resistance ratio (*RRR*) from ~5 to ~30. Magnetization and resistivity measurements show the annealed samples have a sharp ferromagnetic transition with a Curie temperature $T_C$ is 2.8 K. The ordered moment of 0.06 $\mu_B$ is directed along the orthorhombic *c*-axis. Superconductivity is found below a resistive transition temperature $T_s = 0.65$ K.

[*]*Correspondence to*:
Hanoi Advanced School of Science and Technology, Hanoi University of Technology
40 Ta Quang Buu, Hanoi, Vietnam
Tel.: +84432153091; fax: +84436320293
E-mail address: huynt-hast@mail.hut.edu.vn (N.T.Huy)

## 1. Introduction

Recently, we reported a new ambient-pressure ferromagnetic superconductor: UCoGe [1]. The other members of the family of ferromagnetic superconductors are UGe$_2$ [2] (under pressure), UIr [3] (under pressure) and URhGe [4]. Upon cooling these materials to low temperatures, they undergo a transition to an itinerant ferromagnetic state, followed by a superconducting transition in the subkelvin temperature range. Surprisingly, superconductivity coexists with ferromagnetism. The coexistence of superconductivity and ferromagnetism is unusual, because in the standard BCS scenario both ordering phenomena are mutually exclusive. Evidence has been presented that in these ferromagnetic superconductors an alternative scenario is at play: on the border of ferromagnetism, critical ferromagnetic spin fluctuations mediate unconventional superconductivity with spin-triplet pairing [2,5]. However, a generic, microscopic explanation is still lacking.

Ferromagnetism and superconductivity in UCoGe was first reported for polycrystalline samples [1]. UCoGe crystallizes in the orthorhombic TiNiSi structure (space group $P_{nma}$) [6]. Powder X-ray diffraction patterns taken at room temperature confirmed the TiNiSi structure and the lattice parameters extracted are: $a$ = 6.845 Å, $b$ = 4.206 Å and $c$ = 7.222 Å. Magnetization measurements show UCoGe is a weak itinerant ferromagnet with a Curie temperature $T_C$ = 3 K, and a very small ordered moment $m_0$ = 0.03 $\mu_B$. Superconductivity was observed with a resistive transition temperature $T_s$ = 0.8 K for the best samples (as characterized by the residual resistance ratio $RRR = R(300K)/R(1K)$ = 30). The proximity to a ferromagnetic instability, the defect sensitivity of $T_s$, and the absence of Pauli limiting provided evidence that UCoGe is a spin-triplet superconductor [1]. Obviously, in order to explore magnetically mediated superconductivity in UCoGe, high-quality single-crystals are required.

Recently, we have succeeded in preparing single-crystalline samples of UCoGe with good magnetic and superconducting properties. These first crystals enabled us to study the anisotropy in the normal- and superconducting state properties [7]. Magnetization data revealed ferromagnetism in UCoGe is uniaxial and the ordered moment points along the orthorhombic $c$-axis. A study of the anisotropy in the upper critical field $B_{c2}$ provided evidence that the superconducting gap function is unconventional and has point nodes along the direction of $m_0$ [7]. In this paper we describe the sample preparation process. Notably, we show that a long-term annealing procedure can improve the sample quality considerably, *i.e.* from $RRR$ = 5 for the as-grown sample to $RRR$ = 30 for the annealed samples.

2. Experimental

A polycrystalline batch with nominal composition $U_{1.01}CoGe$ was prepared by arc melting the constituents (natural U 3N, Co 3N and Ge 5N) in a water-cooled copper crucible under a high-purity argon atmosphere. Next, a single-crystalline rod was pulled from the melt using a modified Czochralski technique in a tri-arc furnace under a high-purity argon atmosphere. The phase homogeneity and the stoichiometry of samples were investigated by electron probe microanalysis (EPMA). The single crystallinity was checked by Laue backscattering. To improve the sample quality, several pieces of the single crystalline rod, cut by spark-erosion, were wrapped in tantalum foil and annealed in an evacuated quartz tube for one day at 1250 ºC and 21 days at 880 ºC. This heat-treatment procedure is similar to the one applied to the ferromagnetic superconductor URhGe [8,9].

The temperature dependence of the dc magnetization for magnetic fields applied along the different crystallographic directions was measured in a SQUID magnetometer down to temperatures of 2 K and magnetic fields up to 5 T. Measurements of the electrical resistivity were carried out using a standard four-probe low-frequency ac-technique with the current applied along the principal axes in the temperature range between 0.25 and 10 K.

### 3. Results and Discussion

*3.1. As-grown single crystal*

Fig. 1 shows a typical Laue picture (top frame) of the as-grown single crystal UCoGe, which is 4-5 mm in diameter and 40 mm in length (bottom frame). Laue pictures taken at various spots on the sample confirm its single-crystalline nature. EPMA micrographs confirmed the single-phase nature of the crystal. The chemical composition was determined at $U_{0.989}Co_{1.012}Ge_{0.997}$, which is equal to 1:1:1 within the experimental uncertainty of ± 1.5 %.

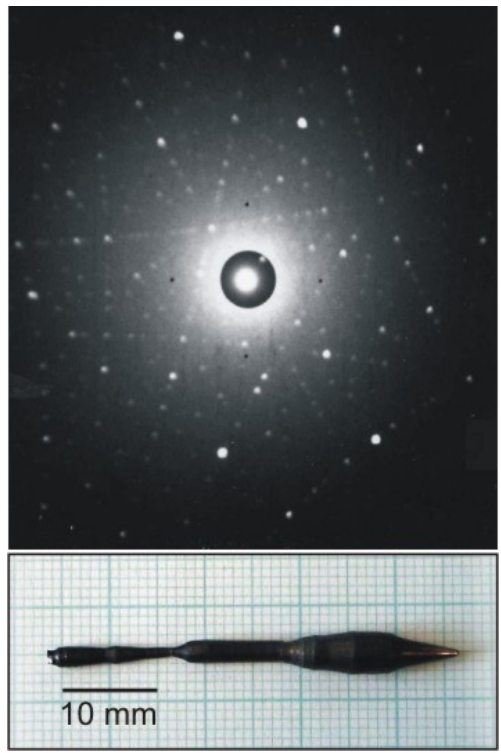

*Fig. 1*. X-ray Laue picture (top) of the as-grown single crystal of UCoGe (photo in lower frame).

The temperature dependence of the magnetization $M(T)$ (in $B$ = 0.01 T) and the resistivity $\rho(T)$ of the as-grown single-crystal UCoGe are presented in Fig. 2. The data show the sample has poor ferromagnetic and superconducting properties. The magnetization (upper panel) is anisotropic. For $B \parallel c$ $M(T)$ increases upon lowering the temperature, while for $B \parallel a, b$ the $M(T)$ curves are flat. The spontaneous magnetization $M_0 \approx 3\times10^{-3}$ $\mu_B$/f.u. at 2 K is much lower than the value $4\times10^{-2}$ $\mu_B$/f.u. at 2 K obtained for the polycrystalline samples [1]. The data for $B \parallel c$ suggest the sample may have a magnetic transition with a Curie temperature below 2 K. In the lower panel of Fig. 2 we show the resistivity data. The onset of superconductivity is observed at 0.33 K, but the transition is not complete at the base temperature of the $^3$He refrigerator (0.24 K). The broad weak hump observed in $\rho(T)$ indicates the ferromagnetic transition is very much smeared. This we attribute to disorder. The residual resistance ratio amounts to ~ 5, which indicates a poor sample quality. The disorder is possibly caused by Co and Ge site inversion. Notice, the TiNiSi structure is an ordered variant of the CeCu$_2$ structure in which Co and Ge atoms randomly occupy the 4c positions [10].

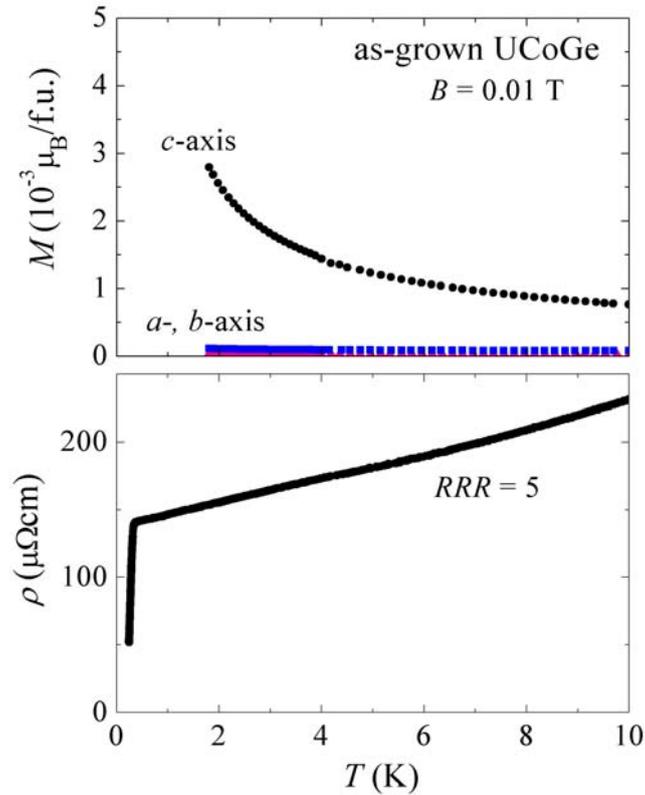

Fig. 2. Magnetization as a function of temperature in a field $B$ of 0.01 T applied along the principal axes of as-grown single-crystalline UCoGe as indicated (*upper panel*). Temperature dependence of the resistivity of the as-grown single-crystal (*lower panel*) with $RRR$ ~ 5. (As the crystal was not oriented, the direction of the current is not specified).

*3.2. Annealed single crystal*

After annealing the single-crystalline nature of the samples is preserved as shown by Laue pictures and, most importantly, the sample quality is significantly improved.

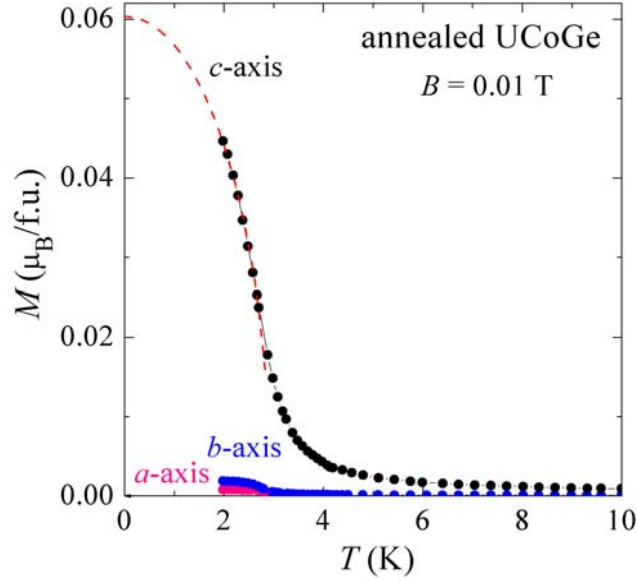

Fig. 3. Magnetization as a function of temperature in a field $B$ of 0.01 T applied along the principal axes of annealed single-crystalline UCoGe as indicated. The dashed line represents a fit to the relation $M(T)^2 = M_0^2(1 - (T/T^*)^2)$ (see text).

The temperature dependence of the magnetization $M(T)$ of annealed single crystalline UCoGe measured in a field of 0.01 T applied along the principal crystallographic axes is shown in Fig. 3. The Curie temperature $T_C$ = 2.8 K is determined by taking the inflection point in $M(T)$ for $B \parallel c$ or the temperature at which $dM(T)/dT$ has a minimum. $T_C$ determined in this way agrees well with $T_C$ deduced from the Arrott plots (not shown) and the location of the sharp kink in the resistivity curve (see Fig. 4). The magnetization curve $M(T)$ for $B \parallel c$ can be described by the relation $M(T)^2 = M_0^2(1 - (T/T^*)^2)$ for weak itinerant ferromagnets [11,12], with the ordered moment $M_0$ = 0.06 $\mu_B$/f.u. and $T^* \sim T_C$. For $B \parallel a, b$ the magnetization values are much lower than for $B \parallel c$. The value $M_0$ = 0.06 $\mu_B$/f.u. obtained from $M(T)$ for $B \parallel c$ is in agreement with the powder averaged value $M_{0,\text{powder}}$ = 0.03 $\mu_B$/f.u. = ½$M_{0,c\text{-axis}}$. The anisotropy in the magnetization data show UCoGe is a uniaxial ferromagnet with the ordered moment pointing along the *c*-axis.

The temperature dependence of the resistivity of annealed single-crystalline UCoGe for a current along the *a*-axis is shown in Fig. 4. The *RRR* value amounts to 30. Proper superconducting and ferromagnetic phase transitions are observed. The magnetic transition is represented by a sharp kink at $T_C$ = 2.8 K, and superconductivity appears at temperatures below $T_s$ = 0.65 K. However, the superconducting transition is still relatively wide, $\Delta T_s$ = 0.1 K. These transition temperatures are in good agreement with those obtained on the best polycrystalline samples [1]. In the ferromagnetic phase ($T_s < T < 0.8T_C$), the resistivity obeys the relation $\rho \sim T^2$ due to scattering at magnons. In the temperature range $T_C < T < 3T_C$ the resistivity is well described by a term $\rho \sim T^{5/3}$ that signals scattering at critical ferromagnetic spin fluctuations [12]. The temperature dependence of the resistivity of UCoGe is characteristic for a weak itinerant electron ferromagnet [12]. The $\rho(T)$ data indicate that superconductivity coexists with ferromagnetism, while the critical scattering provides evidence that UCoGe is located near the border of long-range magnetic order.

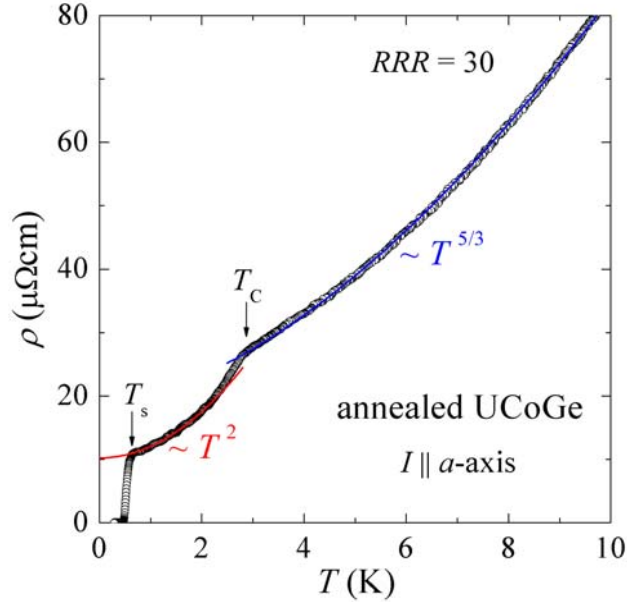

Fig. 4. Temperature variation of the resistivity of annealed single-crystalline UCoGe for a current $I \parallel a$. Arrows indicate the Curie temperature $T_C$ = 2.8 K and the onset temperature for the superconducting transition $T_{s,\text{onset}}$ = 0.65 K. The solid lines represent fits of the data to $\rho \sim T^2$ and $\sim T^{5/3}$ for the temperature ranges below and above $T_C$, respectively. Here the residual resistivity $\rho_0$ = 10.2 μΩcm and $RRR$ = 30.

## 4. Summary

We have reported the effect of annealing on the magnetic and transport properties of single-crystals of the new ferromagnetic superconductor UCoGe. We find that the sample quality is significantly improved by a long-term annealing procedure. The annealed single-crystals order ferromagnetically at $T_C$ = 2.8 K, and superconductivity is observed below $T_s$ = 0.65 K. Magnetization data reveal that UCoGe is a uniaxial ferromagnet with an order moment $M_0$ = 0.06 $\mu_B$ pointing along the orthorhombic *c*-axis. The successful growth of single crystals of UCoGe offers a unique opportunity to study superconductivity stimulated by critical spin fluctuations by techniques like inelastic neutron scattering. Moreover, it offers excellent prospects to probe the intriguing properties of ferromagnetic superconductor in great detail.

**Acknowledgements**

**References**

[1] N. T. Huy, A. Gasparini, D. E. de Nijs, Y. Huang, J. C. P. Klaasse, T. Gortenmulder, A. de Visser, A. Hamann, T. Görlach and H. v. Löhneysen, Phys. Rev. Lett. **99**, 067006 (2007).

[2] S. S. Saxena, K. Ahilan, P.Agarwal, F. M. Grosche, R. K. Haselwimmer, M. Steiner, E. Pugh, I. R. Walker, S. R. Julian, P. Monthoux, G. G. Lonzarich, A. Huxley, I. Sheikin, D. Braithwaite and J. Flouquet, Nature (London) **406**, 587 (2000).

[3] T. Akazawa, H. Hidaka, T. Fujiwara, T. C. Kobayashi, E. Yamamoto, Y. Haga, R. Settai and Y. Ōnuki, J. Phys. Condens. Matter **16**, L29 (2004).

[4] D. Aoki, A. D. Huxley, E. Ressouche, D. Braithwaite, J. Flouquet, J. P. Brison, E. Lhotel and C. Paulsen, Nature **413**, (2001) 613.

[5] D. Fay and J. Appel, Phys. Rev. B **22**, 3173 (1980).

[6] F. Canepa, P. Manfrinetti, M. Pani and A. Palenzona, J. Alloys and Compd. **234**, 225 (1996).

[7] N. T. Huy, D. E. de Nijs, Y. K. Huang and A. de Visser, Phys. Rev. Lett. **100**, 077002 (2008).

[8] D. Aoki, A. Huxley, F. Hardy, D. Braitwaite, E. Ressouche, J. Flouquet, J.P. Brison and C. Paulsen, Acta Phys. Pol. **34**, 503 (2003)

F. Hardy, *Ph.D. Thesis*, University Joseph Fourier, Grenoble (2004), unpublished.

[9] E. Yamamoto, Y. Haga, T.D. Matsuda, Y. Inada, R. Settai, Y. Tokiwa and Y. Onuki, Acta Phys. Pol. **34**, 1059 (2003)

[10] V. Sechovský and L. Havela, *Handbook of Magnetic Materials*, vol. 11, p. 1 (North Holland, Amsterdam, 1998).

[11] G. G. Lonzarich and L. Taillefer, J. Phys. C: Solid State Phys. **18**, 4339 (1985).

[12] T. Moriya, *Spin Fluctuations in Itinerant Electron Magnetism* (Springer, Berlin, 1985).